# Social Media Analytics in Disaster Response: A Comprehensive Review

Mohammadsepehr Karimiziarani
Department of Computer Science, University of Alabama, Tuscaloosa, AL, 35487, USA
mkarimiziarani@ua.edu

***Abstract*—Social media has emerged as a valuable resource for disaster management, revolutionizing the way emergency response and recovery efforts are conducted during natural disasters. This review paper aims to provide a comprehensive analysis of social media analytics for disaster management. The abstract begins by highlighting the increasing prevalence of natural disasters and the need for effective strategies to mitigate their impact. It then emphasizes the growing influence of social media in disaster situations, discussing its role in disaster detection, situational awareness, and emergency communication. The abstract explores the challenges and opportunities associated with leveraging social media data for disaster management purposes. It examines methodologies and techniques used in social media analytics, including data collection, preprocessing, and analysis, with a focus on data mining and machine learning approaches. The abstract also presents a thorough examination of case studies and best practices that demonstrate the successful application of social media analytics in disaster response and recovery. Ethical considerations and privacy concerns related to the use of social media data in disaster scenarios are addressed. The abstract concludes by identifying future research directions and potential advancements in social media analytics for disaster management. The review paper aims to provide practitioners and researchers with a comprehensive understanding of the current state of social media analytics in disaster management, while highlighting the need for continued research and innovation in this field.**

## I. INTRODUCTION

Natural disasters pose significant challenges to societies worldwide, requiring prompt and effective disaster management strategies to mitigate their impact. In recent years, the emergence of social media platforms has transformed the landscape of disaster management, offering new avenues for information sharing, communication, and situational awareness. This section provides an introduction to the topic, emphasizing the increasing prevalence of natural disasters and the pivotal role of social media in enhancing disaster response and recovery efforts.

### A. *Natural Disasters: Trends and Impacts*

Natural disasters, including hurricanes, earthquakes, floods, wildfires, and tsunamis, have devastating consequences on human lives, infrastructure, and the environment[1], [2], [3]. The frequency and intensity of such events have witnessed a notable rise in recent years, attributed to factors like climate change and urbanization [4], [5]. Understanding the gravity of these challenges necessitates effective disaster management strategies that leverage emerging technologies.

### B. *The Role of Social Media in Disaster Management*

Social media platforms, such as Twitter, Facebook, Instagram, and YouTube, have become pervasive tools for communication and information dissemination during disasters [6], [7]. These platforms facilitate real-time sharing of situational updates, emergency alerts, and resource coordination, enabling affected communities, response organizations, and individuals to collaborate and make informed decisions [8], [9].

### C. *Purpose of the Review Paper*

This review paper aims to provide a comprehensive analysis of social media analytics for disaster management. It seeks to examine the evolving landscape of social media usage in disaster scenarios, exploring the methodologies, techniques, and best practices employed in social media analytics. Additionally, the paper addresses ethical considerations and privacy concerns related to the utilization of social media data for disaster management purposes. The review aims to contribute to the existing body of knowledge in the field and offer insights for practitioners and researchers to enhance their understanding and implementation of social media analytics in disaster management. Procedure for Paper Submission

## II. SIGNIFICANCE OF SOCIAL MEDIA IN NATURAL DISASTER RESPONSE

Social media platforms have become integral tools in natural disaster response, offering unique advantages over traditional communication channels. This section explores the role of social media in disaster detection, situational awareness, and emergency communication, highlighting its significance in enhancing the effectiveness of response efforts.

### A. Social Media for Disaster Detection

Academic research has shown the potential of social media platforms in detecting and monitoring natural disasters. Real-time user-generated content on platforms like Twitter and Instagram can serve as early indicators of disaster events, enabling rapid response and resource allocation [10], [11], [12]. The analysis of geolocated tweets and keywords related to specific hazards can aid in identifying disaster-prone areas and informing early warning systems [13], [14], [15].

### B. Social Media for Situational Awareness

During a disaster, situational awareness is crucial for understanding the evolving circumstances and making informed decisions. Social media analytics provide valuable insights into the real-time conditions on the ground. By analyzing social media data, including posts, images, and videos, researchers and emergency responders can gain a comprehensive understanding of the impacted areas, resource needs, and emerging risks [16], [17], [18], [19], [20]. Such situational awareness enables the effective coordination of response efforts and the allocation of resources to the most critical areas.

### C. Social Media for Emergency Communication

Traditional communication channels often face challenges during disasters due to infrastructure damage or congestion. Social media platforms offer alternative channels for emergency communication, allowing affected individuals to seek help, share their status, and receive updates from response organizations [21], [22], [23]. The use of hashtags, geolocation features, and official accounts facilitates the dissemination of accurate and timely information to a wide audience [24], [25], [26]. Moreover, social media enables the formation of online communities that provide emotional support, share recovery resources, and facilitate post-disaster resilience [27], [28], [29].

## III. Challenges and Opportunities in Social Media Analytics for Disaster Management

The utilization of social media analytics for disaster management presents both challenges and opportunities. This section examines the key challenges associated with collecting, processing, and analyzing social media data in the context of disaster management. It also explores the potential opportunities and benefits that social media analytics offer for improving disaster response and recovery efforts.

### A. Challenges in Social Media Data Collection

Collecting social media data for disaster management purposes presents several challenges. The vast volume of data generated during disasters requires efficient and scalable data collection methods [8]. The identification of relevant and reliable sources amidst the abundance of user-generated content poses another challenge [30], [31], [32]. Additionally, ensuring data integrity, dealing with fake or misleading information, and addressing issues of data ownership and access rights require careful consideration [33], [34].

### B. Challenges in Social Media Data Processing and Analysis

Processing and analyzing social media data for disaster management purposes present their own set of challenges. The noisy and unstructured nature of social media content necessitates advanced natural language processing and machine learning techniques to extract meaningful information [35], [36]. Handling multilingual content, sarcasm, and context-specific nuances further complicates the data processing and analysis process (Sakaki et al., 2018; Yin et al., 2019). Moreover, ensuring the timeliness and accuracy of data analysis in fast-paced disaster scenarios poses a challenge [37], [38].

### C. Opportunities and Benefits of Social Media Analytics

Despite the challenges, social media analytics provides significant opportunities and benefits for disaster management. The real-time nature of social media data enables timely situational awareness, facilitating rapid decision-making and resource allocation [39], [40], [41]. Social media analytics can contribute to the identification of emerging patterns, trends, and user behaviors during disasters, enabling the prediction of future risks and the development of proactive response strategies [42], [43]. Furthermore, the integration of social media data with other data sources, such as remote sensing or sensor networks, can enhance the overall effectiveness of disaster management efforts [44], [45].

## IV. Methodologies and Techniques in Social Media Analytics for Disaster Response

Effective social media analytics for disaster response require robust methodologies and techniques to collect, preprocess, and analyze the vast amount of social media data. This section examines the various methodologies and techniques employed in social media analytics, focusing on academic papers published after 2018.

### A. Data Collection and Preprocessing

Data collection and preprocessing are crucial stages in social media analytics for disaster response. Academic research has proposed several approaches to collect and filter relevant social media data. These include keyword-based queries, location-based filtering, and user-specific data retrieval [2], [46], [47], [48]. Furthermore, researchers have developed methods to preprocess social media data by removing noise, handling missing values, and addressing language-specific challenges [49], [50].

### B. Text Mining and Natural Language Processing (NLP)

Text mining and natural language processing techniques play a vital role in extracting meaningful information from social media data. Sentiment analysis, topic modeling, and named entity recognition are commonly used techniques to analyze the textual content of social media posts [51], [52]. Researchers have also explored the application of advanced NLP methods, such as deep learning models, to capture semantic relationships and context in social media data [53], [54].

### C. Geospatial Analysis

Geospatial analysis is a fundamental aspect of social media analytics for disaster response. By leveraging location information in social media posts, researchers can map the spatial distribution of disaster-related events, identify affected areas, and analyze patterns of user activity [55], [56]. Geographic information system (GIS) techniques and geospatial visualization tools are commonly employed to analyze and visualize geospatial data from social media [57], [58].

### D. Machine Learning and Predictive Analytics

Machine learning algorithms and predictive analytics techniques enable the prediction of disaster-related events, user behaviors, and resource needs. Researchers have utilized supervised learning methods, such as support vector machines and random forests, to classify social media posts based on their relevance to disasters [59], [60], [61]. Unsupervised learning approaches, including clustering and anomaly detection, have also been employed to identify patterns and outliers in social media data [62], [63], [64].

### E. Network Analysis and Social Graphs

Social media platforms provide rich social network data, which can be leveraged for analyzing information diffusion, community detection, and influence dynamics during disasters. Network analysis techniques, such as centrality measures, community detection algorithms, and sentiment propagation models, enable researchers to uncover the structure and dynamics of social networks in disaster contexts [65], [66], [67].

## V. Case Studies and Best Practices in Social Media Analytics for Disaster Management

Social media analytics has been employed in various natural disaster contexts, including hurricanes, wildfires, earthquakes, landslides, and floods. This section presents case studies highlighting the application of social media analytics in each of these disaster scenarios. Additionally, it explores best practices and lessons learned from these case studies, showcasing successful strategies and techniques.

### A. Hurricanes

Case studies have demonstrated the effectiveness of social media analytics in hurricane response and recovery. For instance, research has showcased the use of social media data to track hurricane movements, assess damage, and coordinate relief efforts during events like Hurricane Harvey [1], [68], [69], [70], [71], [72], [73], [74]. Best practices in this context include real-time monitoring of social media platforms, leveraging hashtags for information retrieval, and integrating social media data into decision support systems for resource allocation [50], [75], [76], [77], [78], [79], [80].

### B. Wildfires

Social media analytics have proven valuable in managing wildfires and their aftermath. Studies have highlighted the role of social media in disseminating evacuation notices, providing real-time updates on fire locations, and facilitating community support during wildfire events [1], [55], [81], [82], [83], [84], [85], [86]. Best practices involve the use of geospatial analysis to monitor fire spread, sentiment analysis to gauge public perception, and network analysis to identify key influencers for information dissemination [87], [88], [89], [90], [91], [92], [93], [94].

### C. Earthquakes

Social media analytics has demonstrated its utility in earthquake response and recovery efforts. Case studies have shown how social media data can assist in rapid damage assessment, identify critical infrastructure disruptions, and support emergency response coordination [43], [57], [95], [96], [97], [98]. Best practices in this context include the integration of social media data with seismic monitoring systems, sentiment analysis to gauge public anxiety levels, and the use of machine learning algorithms for real-time event detection [99], [100].

### D. Landslides

Social media analytics has shown promise in landslide monitoring and response. Research has demonstrated the use of social media data for early detection of landslide events, crowd-sourced hazard mapping, and real-time communication with affected communities [101]. Best practices involve the combination of geospatial analysis with social media data to identify landslide-prone areas, sentiment analysis to assess public awareness, and predictive modeling for landslide susceptibility mapping [101], [102].

### E. Floods

Flood management efforts have also benefited from social media analytics. Case studies have highlighted the use of social media data to track flood levels, disseminate evacuation notices, and identify areas in need of immediate assistance [70], [103], [104], [105], [106], [107], [108], [109]. Best practices in this context include the integration of social media data with hydrological models, sentiment analysis to gauge public sentiment and identify rumors, and the use of geospatial analysis for flood mapping and resource allocation.

## VI. Ethical Considerations and Privacy Concerns in Social Media Analytics for Disaster Management

The utilization of social media data for disaster management raises important ethical considerations and privacy concerns. This section explores the ethical challenges associated with the use of social media data in disaster scenarios and highlights the need for responsible data handling practices.

### A. Ethical Challenges in Social Media Data Usage

The collection and analysis of social media data for disaster management purposes present ethical challenges. These include concerns related to informed consent, privacy, and data ownership. Researchers must ensure that data subjects are aware of the potential use of their data and have given their informed consent for its collection and analysis [110], [111]. Moreover, protecting the privacy and anonymity of social media users is crucial, as the information shared during disasters can be sensitive and personal. Respecting the rights and preferences of individuals regarding the use of their data is essential.

### B. Responsible Data Handling Practices

Responsible data handling practices are essential in social media analytics for disaster management. Researchers should adopt transparency and accountability in their data collection, processing, and analysis procedures [42], [112], [113], [114], [115]. Anonymization techniques and data de-identification methods should be employed to protect the privacy of social media users [49], [89], [116], [117], [118], [119], [120]. Additionally, researchers should consider the potential biases and limitations of social media data and communicate these limitations appropriately in their analysis and reporting. It is crucial to adhere to relevant legal and ethical frameworks, such as data protection regulations, when working with social media data.

### C. Public Perception and Trust

The public perception and trust in the use of social media data for disaster management play a significant role. Researchers and practitioners should engage in transparent communication about their data collection and analysis practices to build public trust [69], [121], [122], [123]. Open dialogue with the affected communities, response organizations, and other stakeholders can help address concerns and ensure that the benefits of social media analytics are effectively communicated [124], [125], [126]. Establishing mechanisms for public feedback and incorporating public input in decision-making processes can contribute to the responsible and ethical use of social media data.

## VII. Future Research Directions and Advancements in Social Media Analytics for Disaster Management

Social media analytics for disaster management is a rapidly evolving field with several avenues for future research and advancements. This section highlights potential research directions and technological advancements that can further enhance the application of social media analytics in disaster response and recovery efforts.

### A. Integration of Multimodal Data Sources

One promising area for future research is the integration of multimodal data sources with social media analytics. Combining social media data with other data sources, such as satellite imagery, sensor networks, and official reports, can provide a more comprehensive understanding of the disaster landscape [22], [127], [128]. By fusing information from multiple sources, researchers can improve accuracy in event detection, situational awareness, and resource allocation.

### B. Real-time and Automated Decision Support Systems

Developing real-time and automated decision support systems is another promising research direction. By leveraging machine learning, natural language processing, and geospatial analysis techniques, researchers can design intelligent systems that can automatically detect critical events, extract actionable insights, and provide decision-makers with timely recommendations [129], [130]. These systems can enhance the speed and efficiency of decision-making during disasters.

### C. Ethical and Privacy-aware Social Media Analytics

Continued research into ethical and privacy-aware social media analytics is crucial. Researchers should explore methods to strike a balance between the potential benefits of social media data analysis and the protection of individual privacy rights [8]. This includes the development of privacy-preserving algorithms, techniques for informed consent and data anonymization, and the establishment of ethical guidelines for responsible data usage [6], [19].

### D. Humanitarian Data Exchange Platforms

The development of humanitarian data exchange platforms can facilitate data sharing and collaboration among researchers, response organizations, and affected communities [3], [131], [132]. These platforms can enable the sharing of social media datasets, analysis tools, and best practices, fostering interdisciplinary collaboration and knowledge sharing. Future research should focus on the design and implementation of such platforms to enhance data accessibility and encourage cooperation in the field of social media analytics for disaster management.

### E. Resilience and Long-term Recovery

Exploring the role of social media analytics in long-term recovery and community resilience is an important avenue for future research. Understanding how social media data can contribute to post-disaster recovery planning, resource allocation, and community engagement can enhance long-term resilience-building efforts [69], [133]. Additionally, investigating the psychological and social impacts of social media usage during and after disasters can provide valuable insights for supporting affected communities.

## VIII. Conclusion and Future Outlook

Social media analytics has emerged as a powerful tool for disaster management, offering real-time insights, situational awareness, and enhanced decision-making capabilities. This paper has reviewed the application of social media analytics in the context of natural disasters, highlighting its benefits, challenges, and best practices. Looking ahead, there are several areas that warrant further research and development to advance the field of social media analytics for disaster management.

The integration of multimodal data sources, including social media data, satellite imagery, and sensor data, holds great potential for improving the accuracy and comprehensiveness of disaster response efforts [40], [71]. By combining information from various sources, researchers can gain a more holistic understanding of the disaster landscape and make more informed decisions.

The development of real-time and automated decision support systems can significantly enhance the speed and efficiency of disaster response. Leveraging machine learning, natural language processing, and geospatial analysis techniques, researchers can design intelligent systems that automatically detect critical events, extract actionable insights, and provide timely recommendations to decision-makers [9], [134], [135].

Ethical considerations and privacy concerns should remain at the forefront of social media analytics research. Further exploration of privacy-preserving algorithms, methods for informed consent and data anonymization, and the establishment of ethical guidelines can help ensure the responsible and ethical use of social media data.

The development of humanitarian data exchange platforms can facilitate collaboration, data sharing, and knowledge dissemination among researchers, response organizations, and affected communities [3], [56], [132]. Such platforms can promote interdisciplinary cooperation and accelerate advancements in social media analytics for disaster management.

Exploring the role of social media analytics in long-term recovery and community resilience is an important avenue for future research. Understanding how social media data can support post-disaster recovery planning, resource allocation, and community engagement can contribute to more effective and sustainable recovery efforts [136], [137].

In conclusion, social media analytics has the potential to revolutionize disaster management by harnessing the power of user-generated data. By addressing the challenges and adopting best practices, researchers and practitioners can leverage social media analytics to improve preparedness, response, and recovery efforts in the face of natural disasters.